\documentstyle[multicol,aps]{revtex} 
\begin{document}
\draft
 
\title{Complete Exact Solution of Diffusion-Limited Coalescence,
$A+A\to A$}
\author{Daniel ben-Avraham\footnote
  {{\bf e-mail:} qd00@polaris.clarkson.edu}}

\address{Physics Department, and Clarkson Institute for Statistical
Physics (CISP), \\ Clarkson University, Potsdam, NY 13699-5820}
\maketitle

\begin{abstract} 
Some models of diffusion-limited reaction processes in one dimension lend
themselves to exact analysis.  The known approaches yield exact
expressions for a limited number of quantities of interest, such as the particle
concentration, or the distribution of distances between nearest particles. 
However, a full characterization of a particle system is only provided by the
infinite hierarchy of multiple-point density correlation functions.  We derive
an exact description of the full hierarchy of correlation functions for the
diffusion-limited irreversible coalescence process $A+A\to A$. 
\end{abstract}

Non-equilibrium kinetics of diffusion-limited reactions has been the subject
of much recent interest
\cite{vanKampen,Haken,Nicolis,Ligget,reaction-reviews,JStatPhys}.  In
contrast to equilibrium systems --- which are best analyzed with standard
thermodynamics --- or reaction-limited processes --- whose kinetics is well
described by classical rate equations
\cite{Laidler,Benson} --- there is no general approach to non-equilibrium,
diffusion-limited reactions.  Some diffusion-limited processes in one
dimension can be approached analytically, but the solution methods yield
only a  limited number of quantities of interest, such as the particle
concentration, or the distribution of distances between nearest particles.
In this letter, we derive an exact analytic description of the full hierarchy
of multiple-point density correlation functions for the
irreversible diffusion-limited coalescence process $A+A\to A$.  The
hierarchy of correlation functions characterizes the system completely.

Our model
\cite{Coalescence,Doering,AAreviews} is defined on the line
$-\infty<x<\infty$.  Particles $A$ are represented by points which perform
unbiased diffusion with a diffusion constant $D$. When two particles meet
they merge into one particle which continues diffusing with the same diffusion
constant $D$ as the reacting particles.  Since the reaction step is infinitely
fast, the system models the {\it diffusion-limited\/} coalescence process
$A+A\to A$.

An exact treatment of the problem is possible through the method of Empty
Intervals, known also as the method of Inter-Particle Distribution Functions
(IPDF).  The key parameter is $E(x,y;t)$ --- the probability that the interval
$[x,y]$ is empty (contains no particles) at time $t$.  Particles just at the
edge of an empty interval may diffuse into or out of the interval, affecting
the probability
$E$.  With this observation in mind, one can write down a rate equation for
the empty interval probability
\cite{Coalescence,Doering}. For the model at hand, the rate equation is
\begin{equation}
\label{dE/dt}
{\partial E(x,y;t)\over\partial t}=D({\partial^2\over\partial x^2}
  +{\partial^2\over\partial y^2})E(x,y;t)\;.
\end{equation}
The coalescence reaction imposes the boundary condition
\begin{equation}
\label{bc:E=1}
\lim_{y\uparrow x{\rm\ or\ }x\downarrow y}E(x,y;t)=1\;,
\end{equation}
and in addition, as long as there are any particles left in the system,
\begin{equation}
\label{bc:E=0}
\lim_{{x\to-\infty\atop y\to+\infty}}E(x,y;t)=0\;.
\end{equation}
{}From $E(x,y;t)$ one can derive useful parameters, such as the concentration
of particles;
\begin{equation}
\label{rho1}
\rho(x;t)=-{\partial\over\partial y}E(x,y;t)|_{y=x}\;,
\end{equation}
or the probability that given a particle at $x$ the next nearest
particle is at $y$ (the IPDF);
\begin{equation}
\label{IPDF}
p(x,y;t)=\rho(x,y;t)^{-1}{\partial^2\over\partial x\,\partial y}E(x,y;t)\;.
\end{equation}

While such level of description affords us invaluable physical insights into
many interesting applications~\cite{AAreviews}, it does not characterize the
system completely.  For that we need the multiple-point density correlation
functions $\rho_n(x_1,x_2,\dots,x_n;t)$, {\it i.e.\/}, the joint probability
density to find $n$ particles at $x_1,x_2,\cdots,x_n$ at time $t$.  For
$n=1$, $\rho_1(x,t)$ is identical with the particle concentration of
Eq.~(\ref{rho1}).  However, a complete characterization of the system
requires knowledge of the {\it full\/} hierarchy
of correlation functions $\{\rho_n\}_{n=1}^{\infty}$.

The multiple-point correlation functions may be obtained from a
generalization of the method of Empty Intervals~\cite{Doering}.  Let
$E_n(x_1,y_1,x_2,y_2,\dots,x_n,y_n;t)$ be the joint probability that the
intervals $[x_i,y_i]$ ($i=1,2,\dots,n)$ are empty at time $t$.  The
intervals are non-overlapping, and ordered: $x_1<y_1<\cdots<x_n<y_n$. Then,
the $n$-point correlation function is given by
\begin{equation}
\label{rho_n}
\rho_n(x_1,\dots,x_n;t)=
(-1)^n{\partial^n\over\partial y_1\cdots\partial y_n}
E_n(x_1,y_1,\dots,x_n,y_n;t)|_{y_1=x_1,\dots,y_n=x_n}\;.
\end{equation}
Doering~\cite{Doering} has shown that for irreversible coalescence the $E_n$
satisfy the partial differential equation:
\begin{equation}
\label{dEn/dt}
{\partial\over\partial t} E_n(x_1,y_1,\dots,x_n,y_n;t)
=D({\partial^2\over\partial x_1^2} +{\partial^2\over\partial y_1^2} + \cdots
+{\partial^2\over\partial x_n^2} + {\partial^2\over\partial y_n^2})E_n\;,
\end{equation}
with the boundary conditions
\begin{equation}
\label{bc:En.1}
\lim_{x_i\uparrow y_i{\rm\ or\ }y_i\downarrow x_i}
  E_n(x_1,y_1,\dots,x_n,y_n;t)=
  E_{n-1}(x_1,y_1,\dots,\not{\!x_i},\not{\!y_i},\dots,x_n,y_n;t)\;,
\end{equation}
and
\begin{equation}
\label{bc:En.2}
\lim_{y_i\uparrow x_{i+1}{\rm\ or\ }x_{i+1}\downarrow y_i}
 E_n(x_1,y_1,\dots,x_n,y_n;t)=
 E_{n-1}(x_1,y_1,\dots,\not{\!y_i},\not{\!x_{i+1}},\dots,x_n,y_n;t)\;.
\end{equation}
For convenience, we use the notation that crossed out arguments (e.g.
${\not{\!x_i}}$) have been removed.  The $E_n$ are tied together
in an hierarchical fashion through the boundary conditions~(\ref{bc:En.1})
and (\ref{bc:En.2}):  one must know $E_{n-1}$ in order to compute $E_n$.

We now provide with a full solution of the $E_n$.  At the level of $n=2$, the
solution is~\cite{unpublished}:
\begin{eqnarray}
\label{E2}
&&E_2(x_1,y_1,x_2,y_2;t)= \nonumber\\
&&E(x_1,y_1;t)E(x_2,y_2;t)-E(x_1,x_2;t)E(y_1,y_2;t)
  +E(x_1,y_2;t)E(y_1,x_2;t)\;.
\end{eqnarray}
Clearly, each term on the r.h.s. of Eq.~(\ref{E2}) satisfies the
equation~(\ref{dEn/dt}), as well as the boundary conditions~(\ref{bc:En.1})
and (\ref{bc:En.2}) --- since $E(x,x;t)=1$
(Eq.~\ref{bc:E=1}). Similarly, $E_n$ can also be expressed in terms of
products of $E$'s for single intervals:
\begin{equation}
\label{En}
E_n(x_1,y_1,\dots,x_n,y_n;t) = 
  \sum_{p=1}^{(2n-1)!!}\sigma_pE(z_{1,p},z_{2,p};t)E(z_{3,p},z_{4,p};t)\cdots 
  E(z_{2n-1,p},z_{2n,p};t)\;.
\end{equation}
Here $z_{1,p},z_{2,p},\dots,z_{2n,p}$ symbolize an {\it ordered\/}
permutation, $p$, of the variables $x_1,y_1,\dots,x_n,y_n$, such that
\begin{equation}
\label{order}
z_{1,p}<z_{2,p},\;z_{3,p}<z_{4,p},\dots,z_{2n-1,p}<z_{2n,p},\quad{\rm and}\quad 
z_{1,p}<z_{3,p}<z_{5,p}\cdots<z_{2n-1,p}\;.
\end{equation}
There are exactly $(2n-1)!!=1\cdot3\cdot\,\cdots\,\cdot(2n-1)$
such permutations. $\sigma_p$ is $+1$ for even permutations
(permutations that require an even number of exchanges between pairs of
variables), or $-1$ for odd permutations.

Eq.~(\ref{En}) can be proved by induction.  For $n=1$ it reduces
to $E_1(x,y;t)=E(x,y;t)$, and for $n=2$ it reduces to Eq.~(\ref{E2}), as
required.  We need only show that if~(\ref{En}) is true for $n-1$ ($n\geq
3$), then it is also valid for $n$. It is easy to see that because each of the
$E$'s satisfies Eq.~(\ref{dE/dt}), the proposed $E_n$ satisfies
Eq.~(\ref{dEn/dt}).  Now test the boundary conditions: Suppose that
$x_i=y_i$.  The permutations in the r.h.s of~(\ref{En}) are divided into
two groups: (a) those which leave $(x_i,y_i)$ as an argument of one of the
$E$'s in the product, and (b) those which separate $x_i$ and $y_i$ into
different $E$'s in the product.  The permutations in group (a) add up to
$E_{n-1}(x_1,y_1,\dots,{\not{\!x_i}},{\not{\!y_i}},\dots,x_n,y_n;t)$, as
required by the boundary condition~(\ref{bc:En.1}).  This is because
$E(x_i,y_i;)=1$ (from Eq.~\ref{bc:E=1}), and because of the induction
assumption regarding the validity of~(\ref{En}) for $n-1$.  Notice that the
parity of the permutations in group (a) is the same as if $x_i$ and $y_i$ were
removed. On the other hand, the permutations in group (b) add up to zero, for
the following reason. Suppose that in some permutation $x_i$ and $y_i$ are
paired with other variables $z_1$ and $z_2$; $(x_i,z_1)$ and $(y_i,z_2)$. 
Then, there exists a similar permutation of the variables
$x_1,y_1,\cdots,x_n,y_n$ where individual pairs remain in the same order, but
now the pairings of
$x_i$ and $y_i$ are exchanged; $(x_i,z_2)$ and $(y_i,z_1)$.  When $x_i=y_i$,
the product of the $E$'s in these two permutations is identical.  But the
parity of the two permutations is opposite, and so they add up to zero.  The
same is true if the pairings of $x_i$ and $y_i$ are
$(z_1,x_i)$ and $(z_2,y_i)$, or $(z_1,x_i)$ and $(y_i,z_2)$.  [Notice that
the pairings $(x_i,z_1)$ and $(z_2,y_i)$ cannot occur, because of the
required ordering, Eq.~(\ref{order}).]  In sum, the boundary
condition~(\ref{bc:En.1}) is satisfied.  The proof of~(\ref{bc:En.2}) follows
a similar line of reasoning.

The
ordered permutations of the endpoints of $n$ intervals ($2n$ variables) may
be constructed {\it recursively\/} in the following way.  The order constraint
of Eq.~(\ref{order}) requires that
$z_{1,p}=x_1$, for all permutations $p$.   Set then
$z_{1}=x_1$, and $z_{2}$ equal to one of the other
$y_1,\dots,x_n,y_n$ variables. Then, arrange the remaining $2n-2$ variables
in all their ordered permutations.  Finally, repeat this procedure, selecting
sequentially
$z_2=y_1,x_2,y_2,\dots,x_n,y_n$.  Thus, the number of
permutations for the endpoints of $n$ intervals,
$N(n)$, satisfies the recursion relation $N(n)=(2n-1)N(n-1)$, and therefore
$N(n)=(2n-1)!!$ --- since $N(1)=1$.  
The recursive construction allows us also to express $E_n$ more compactly, in
terms of $E_{n-1}$:
\begin{eqnarray}
\label{En.recursive}
E_n(x_1,y_1,\dots,x_n,y_n;t) =&& 
+\sum_{j=1}^nE(x_1,y_j;t)E_{n-1}(\not{\!x_1},y_1,\dots,x_j,\not{\!y_j},\dots,
  x_n,y_n;t) \nonumber\\
&&-\sum_{j=2}^nE(x_1,x_j;t)E_{n-1}(\not{\!x_1},y_1,\dots,\not{\!x_j},y_j,\dots,
  x_n,y_n;t) \;.
\end{eqnarray}

Until now we have ignored the issue of {\it initial conditions\/}. For
the solution~(\ref{En}) to work, it is required that the same relation be
satisfied at time $t=0$ as well!  This seems at first sight a formidable
restriction, but fortunately some most important situations are unaffected by
it.  If the particles are initially randomly distributed,
independently from each other, at a homogeneous concentration
$\rho_0$, then $E(x,y;0)=e^{-\rho_0(y-x)}$ and
\begin{equation}
E_n(x_1,y_1,\dots,x_n,y_n;0)=e^{-\rho_0[(y_1-x_1)+\cdots+(y_n-x_n)]}\;.
\end{equation}
One can easily check that Eq.~(\ref{En}) is satisfied.  In this case,
Eq.~(\ref{rho_n}) yields the obvious relation:
$\rho(x_1,x_2,\dots,x_n;0)=\rho_0^n$.  This uncorrelated random initial
distribution gets quickly correlated with time.
  
Another interesting situation is the state of the system in the long-time
asymptotic limit.  It can be shown that the system arrives at a {\it
universal\/} asymptotic state, independent of the initial distribution of
particles.  (This excludes some exotic situations, such as fractal initial
distributions.)  The long-time asymptotic solution of
Eq.~(\ref{dE/dt}), with the boundary conditions~(\ref{bc:E=1}) and
(\ref{bc:E=0}) is
\begin{equation}
E(x,y;t)={\rm erfc}\big({y-x\over\sqrt{8Dt}}\big)\;.
\end{equation}
Using Eq.~(\ref{rho1}), the corresponding long-time asymptotic
concentration is:
\begin{equation}
\rho_{\rm asymp.}(x;t)={1\over\sqrt{2\pi Dt}}\;.
\end{equation}
{}From Eqs.~(\ref{En}) and (\ref{rho_n}) we get the two-point correlation
function
\begin{equation}
\label{rho2}
{\rho_2(x_1,x_2;t)\over\rho_{\rm asymp.}^2}=
1-e^{-2\xi^2}+\sqrt{\pi}\,\xi e^{-\xi^2}{\rm erfc}(\xi)\;,
\end{equation}
where we used the notation $\xi=(x_2-x_1)/\sqrt{8Dt}$.   For the three-point
correlation function, we get
\begin{eqnarray}
\label{rho3}
{\rho_3(x_1,x_2,x_3;t)\over\rho_{\rm asymp.}^3} &=&
1-e^{-2\xi_{21}^2}-e^{-2\xi_{32}^2}-e^{-2\xi_{31}^2}
+2e^{-\xi_{21}^2-\xi_{32}^2-\xi_{31}^2}\nonumber\\
&&+\sqrt{\pi}\,\xi_{21} 
     (e^{-\xi_{21}^2}-e^{-\xi_{32}^2-\xi_{31}^2})
                               {\rm erfc}(\xi_{21})\nonumber\\ 
&&+\sqrt{\pi}\,\xi_{32}
     (e^{-\xi_{32}^2}-e^{-\xi_{21}^2-\xi_{31}^2})
                               {\rm erfc}(\xi_{32})\nonumber\\  
&&+\sqrt{\pi}\,\xi_{31}
     (e^{-\xi_{31}^2}-e^{-\xi_{21}^2-\xi_{32}^2}){\rm erfc}(\xi_{31})\;,
\end{eqnarray}
where now $\xi_{ij}=(x_i-x_j)/\sqrt{8Dt}\,$ (notice that
$\xi_{31}=\xi_{32}+\xi_{21}$ is {\it not\/} an independent variable).

In Fig.~\ref{fig1}, we show the two-point correlation function in the
long-time asymptotic limit (Eq.~\ref{rho2}). We see that the two points become
uncorrelated as the distance between them increases, but that there is an
effective strong repulsive interaction (due to the calescence reaction)
between nearby particles.  Interestingly, the two-point correlation is a
monotonous function of the distance.  A simple convolution of the 
distances between nearest particles predicts an oscillating
tail~\cite{Alemany-DbA}.

The three-point
correlation function, Eq.~(\ref{rho3}), is a bit harder to illustrate.
Instead of a full description, in Fig.~\ref{fig2} we compare
$\rho_3(x_1,x_2,x_3)$ to $\rho_2(x_1,x_2)\rho_2(x_2,x_3)/\rho(x_2)$ (in the
spirit of the truncation ansatz that might be used in a Kirkwood
approximation,say), 
along the line $x_3-x_2=x_2-x_1$.  Again, we see that as the distance
between the three particles increases they become rapidly uncorrelated, but
that the approximation ansatz fails for short distances, due to reactions.

In summary, we have obtained the complete hierarchy of $n$-point density
correlation functions for the diffusion-limited irreversible coalescence
process, $A+A\to A$, in one dimension.  The recursive form of
Eq.~(\ref{En.recursive}) should be useful to prove theorems.  Moreover, with
today's available software for symbolic manipulation, it is relatively easy to
churn out explicit expressions for a finite number of points.  In fact, we
have used ``Mathematica" to compute the examples of 2- and 3-point
correlation functions.

The more general
problems of reversible coalescence --- when the back reaction $A\to A+A$ is
allowed --- and coalescence with different kinds of particle input, can also
be handled, in principle, by the empty interval approach.  That is,
in both cases (even both at the same time) there is a closed hierarchy of
linear partial differential equations, coupled through their boundary
conditions, similar to Eqs.~(\ref{dEn/dt}), (\ref{bc:En.1}) and
(\ref{bc:En.2})
\cite{Doering}. But the linear operators involved break the simple $x$-$y$
symmetry of Eq.~(\ref{dEn/dt}), and a complete solution remains an open
challenge.  

\acknowledgments
I thank Larry Glasser for useful discussions, and Charlie Doering for
sharing with me his elegant solution for the two-point
correlation function.


\begin{figure}
\caption{Two-point correlation function for the coalescence process in the
long-time asymptotic limit. Shown is
$\rho_2(\xi)/\rho_{\rm asymp.}^2$ vs. $\xi$.}
\label{fig1}
\end{figure} 

\begin{figure}
\caption{Three-point correlation function for the coalescence process in the
long-time asymptotic limit.  Plotted is the relative error made by the
Kirkwood approximation,
$(\rho_{\rm Kirkwood}-\rho_3)/\rho_3$, for the line
$\xi_{21}=\xi_{32}\equiv\xi$.   The Kirkwood approximation in this case is
$\rho_{\rm Kirkwood}=\rho_2(\xi)^2/\rho_{\rm asymp.}$.} 
\label{fig2}
\end{figure}

\end{document}